\newif\ifsubmission
\submissiontrue

\newif\ifanonymous
\anonymousfalse

\documentclass[conference]{IEEEtran}
\IEEEoverridecommandlockouts
\usepackage{cite}
\usepackage{amsmath,amssymb,amsfonts}
\usepackage{algorithmic}
\usepackage{graphicx}
\usepackage{textcomp}
\usepackage{xcolor}
\usepackage{booktabs}
\usepackage{enumitem}
\usepackage{float}
\usepackage{etoolbox}
\usepackage{lipsum}

\def\BibTeX{{\rm B\kern-.05em{\sc i\kern-.025em b}\kern-.08em
    T\kern-.1667em\lower.7ex\hbox{E}\kern-.125emX}}

\ifsubmission
  \newcommand{\TODO}[1]{}
  
  \newcommand{\remove}[1]{}
\else 
  \newcommand{\TODO}[1]{\textcolor{orange}{ToDo\footnote{#1}}}
  
  \newcommand{\remove}[1]{\textcolor{red}{#1}}
\fi

\newcommand{\sys}[1]{\textsc{SynFi}}

\newcommand{\sect}[1]{Section~\ref{#1}}
\newcommand{\fig}[1]{Figure~\ref{#1}}
\newcommand{\tab}[1]{Table~\ref{#1}}

\renewcommand{\paragraph}[1]{\vspace{0.5em}\noindent\textbf{#1}.}

\newcommand{\para}[1]{\vspace{3pt} \noindent \textbf{#1.}}

\makeatletter
\let\@authorsaddresses\@empty
\makeatother

\ifanonymous
\else
\makeatletter
\patchcmd{\@maketitle}
  {\addvspace{0.5\baselineskip}\egroup}
  {\addvspace{-1\baselineskip}\egroup}
  {}
  {}
\makeatother
\fi

\newcommand\blfootnote[1]{%
  \begingroup
  \renewcommand\thefootnote{}\footnote{#1}%
  \addtocounter{footnote}{-1}%
  \endgroup
}
    
\begin{document}

\title{{\sys{}: Automatic Synthetic Fingerprint Generation}
\ifanonymous
\else
\fi
}

\ifanonymous
\author{}
\else
\author{
\IEEEauthorblockN{M. Sadegh Riazi$^*$}
\IEEEauthorblockA{
\textit{UC San Diego}\\
mriazi@ucsd.edu}
\and
\IEEEauthorblockN{Seyed M. Chavoshian$^*$}
\IEEEauthorblockA{
\textit{UC San Diego}\\
schavosh@ucsd.edu}
\and
\IEEEauthorblockN{Farinaz Koushanfar}
\IEEEauthorblockA{
\textit{UC San Diego}\\
farinaz@ucsd.edu}
}
\fi

\maketitle

\begin{abstract}
Authentication and identification methods based on human fingerprints are ubiquitous in several systems ranging from government organizations to consumer products.
The performance and reliability of such systems directly rely on the volume of data on which they have been verified.  
Unfortunately, a large volume of fingerprint databases is not publicly available due to many privacy and security concerns. 

In this paper, we introduce a new approach to automatically generate high-fidelity synthetic fingerprints at scale. 
Our approach relies on (i) Generative Adversarial Networks to estimate the probability distribution of human fingerprints and (ii) Super-Resolution methods to synthesize fine-grained textures. 
We rigorously test our system and show that our methodology is the \textit{first} to generate fingerprints that are computationally indistinguishable from real ones, a task that prior art could not accomplish.
\end{abstract}

\ifanonymous
\vspace{0.2cm}
\fi

\graphicspath{{figs/}} \sloppy

\section{Introduction}\label{sec:intro}
Human fingerprints are frequently used in several applications for authentication and identification, ranging from smart doors to authorizing payments on cell phones.  
Evaluating the performance and reliability of identification and verification fingerprint-based systems requires access to a large fingerprint database. 
However, in practice, obtaining a massive corpus of fingerprint images incurs a high cost. 
In many cases, the research groups that are developing fingerprint-based authentication systems, do not have access to a large publicly-available database. 
The performance of these systems is directly dependent on the quality and quantity of the available data.
\ifanonymous
\else
~\blfootnote{$^*$Equal contributions of the first two authors.}
\fi

In addition to the above obstacles, gathering fingerprint impressions of a large population of people raises severe privacy and security concerns. 
In case of a breach, the fingerprint of many users will be directly exposed to attackers and can be used to fool any other authentication systems that accept fingerprints. 
To this end, we study the task of generating \emph{synthetic fingerprints} which can solve the challenges mentioned above.
Synthetic fingerprints solve the availability concern as they can be generated for virtually any number of samples. 
Moreover, synthetic fingerprints are artificially generated; hence, they do not leak any information about real identities. 

Synthetic fingerprints also play essential roles in other tasks as well. 
For example, they can be used to analyze the robustness of a verification system against Trojan attacks~\cite{maltoni2009handbook}.
To perform this analysis, a large number of fingerprints are needed where their fine-grained features can be varied while fixing other characteristics such as image orientation.

Fingerprints can be categorized based on their global structure and curvatures. 
Some of these categories are drastically rarer, and their synthetic counterparts can be used to compensate for the imbalance. 
One can generate a specific type of fingerprints that are rarer in the real-world.
Moreover, the security of several biometric storage mechanisms that protect fingerprints in case of a breach relies on the assumption that the size of a database is bigger than a specific threshold~\cite{chatterjee2019multisketches}. 
In these scenarios, synthetic fingerprints can be used as a means to populate small databases.

Prior synthetic fingerprint generation solutions were able to either create synthetic \emph{templates} of fingerprint's micro-features or synthetic image of actual fingerprints but at a low resolution. 
Solutions based on mathematical models of fingerprints suffer from lack of entropy and generalization to accurate probability distribution of real fingerprints~\cite{cappelli2004sfinge}. 
Prior solutions based on Deep Learning (DL) models also cannot produce high-quality images due to the small volume of the available real fingerprints to train these models~\cite{bontrager2018deepmasterprints}.

In this paper, we present \sys{}, a new comprehensive framework to automatically generate high-quality synthetic fingerprints at scale. 
Our solution formulates the process of generating synthetic fingerprints as two parallel deep learning tasks based on Generative Adversarial Network (GAN) and Super-Resolution (SR) paradigm. 
In particular, \sys{} formalizes and satisfies the following design goals to meet real-world expectations: 
(i) the generated samples should preserve the minutiae characteristics of fingerprints used for authentication systems, e.g., ridge structure, bifurcations, and ridge endings. 
(ii) An ideal system should be able to generate \emph{full-finger impressions} as opposed to partial fingerprints. 
(iii) Synthetic fingerprints should be \emph{computationally indistinguishable} from real impressions to be used as a means to extend the security of biometric storage systems. 
(iv) The system should be fully automated, requiring no manual feature engineering to have high \emph{scalability}.  
As we show in the rest of this paper, \sys{} satisfies all of the above requirements. 

\para{Contributions}
Our concrete contributions are as follows.
\begin{itemize}[leftmargin=*]
    \item We propose a new framework to generate robust full-finger synthetic fingerprints. 
    We explore several deep learning-based solutions to generate high-quality samples. 
    We formulate this task based on generative adversarial network and super-resolution methodologies. 
    \item We perform qualitative as well as quantitative analysis on distinguishability of synthetic fingerprints from real ones using six different machine learning models.
    \item We provide the proof-of-concept implementation of our proposed methodology in Pytorch.
    We open-sourced\footnote{https://github.com/MohammadChavosh/synthetic-fingerprint-generation} our framework to facilitate progress, improvements, and verification process of fingerprint-based systems.
\end{itemize}

\section{Prior Art}
The prior work on generating synthetic fingerprints can be categorized into two broad groups. 
The first group is based on formulating mathematical models to generate artificial fingerprints. 
The second group leverages various classes of deep learning models. 
Generally speaking, the first group involves more feature engineering and manual tuning, whereas the second group inherits a more automated nature of feature extraction of the DL models. 

\para{Mathematical Models}
One of the systems based on mathematical models is called SFinGe~\cite{cappelli2004sfinge}. 
In this system, generating a synthetic fingerprint involves four main phases: 
(i) a fingerprint shape is randomly generated via specific geometric models,
(ii) a directional map is produced,
(iii) a density map is created, and 
(iv) the first three maps are combined to generate a fingerprint pattern using a ridge-flow model.
Finally, noise is added to make the generated image more realistic.

Unfortunately, solutions based on the mathematical models suffer from the low level of entropy due to the rigorous structure of the generation process. 
In contrast, \sys{} generates each fingerprint starting from a completely \emph{random noise}.

\para{Deep Learning Models}
Deep learning has demonstrated a breakthrough in several applications and domains. 
There are several categories of DL models. Two of which that are explored for the task of synthetic fingerprint generation are 
(i) Fully Visible Belief Networks (FVBN)
such as PixelRNN~\cite{oord2016pixel} that can produce one pixel at a time.
Similar to Recurrent Neural Networks (RNN) that generate text, FVBNs can be used to create pixels of a fingerprint image. 
One drawback of these networks is that the final output can often be noisy. 
(ii) The second group is based on Variational Autoencoders (VAE).
Compared to FVBN, VAE usually produces smoother images. 
Another line of work focuses on {\it MasterPrints}, which are real or synthetic fingerprint templates at \emph{feature-level}~\cite{roy2018evolutionary} that can fool fingerprint-based authentication systems and authenticate the attacker as a legitimate user. 
The idea was later generalized to DeepMasterPrints, which are synthetic fingerprints at \emph{image-level}~\cite{bontrager2018deepmasterprints}.

However, state-of-the-art DL-based methods can only generate \emph{low-quality partial} fingerprints. 
In contrast, as we compare in \sect{sec:exp}, \sys{} generates full-finger fingerprints with significantly higher resolution due to a novel DL formulation.

\para{Other Related Work}
Secure Multi-Party Computation (SMPC) protocols can be used to enhance the security and privacy of users' data during identification and authentication process~\cite{riazi2019xonn,riazi2018chameleon,riazi2019deep}. 
Given the fingerprint matching algorithm, SMPC protocols can securely execute the authentication algorithm without exposing user's fingerprint to the authenticating server. 
However, such solutions cannot enhance the reliability and performance of the matching algorithm itself, and thus, are complementary to \sys{}.




\section{Preliminaries}
\para{Fingerprint Features}
Each fingerprint has a set of micro-features that can be used to uniquely identify the finger. 
\emph{Minutiae} points correspond to the particular locations of fingerprint, e.g., ridge bifurcations or endings. 
Each Minutia is represented as a tuple of the location ($x$, $y$), orientation $\theta$, and a quality factor $q$. Since different impressions of the same finger can result in drastically different minutiae tuples, fingerprint matching methodologies rely on a scale- and rotation-invariant algorithms to detect whether two sets of minutiae points belong to the same finger or not. 
An ideal synthetic fingerprint generator should produce impressions such that the distribution of minutiae points is not far from real ones.

\para{Generative Adversarial Networks}
One family of deep learning models that has become very popular in recent years is Generative Adversarial Networks (GAN)~\cite{goodfellow2014generative}. 
Using GANs, one can estimate the distribution of a given dataset. 
The key idea in GANs is to train two different neural networks in parallel and make use of each of them to improve the other one. One of the networks, the \emph{generative} network, tries to generate samples from the given data distribution. The other one, the \emph{discriminator} network, is in charge of learning to distinguish the samples generated by the generative network from the real data samples. 
During the training phase, the feedback from the discriminator network is used to enhance the quality of the samples produced by the generative network.

After their introduction, GANs has been used in different areas and tasks. One of their most important applications is generating synthetic images. 
For example, \cite{karras2019style} generates artificial images of human faces and \cite{park2019semantic} uses GANs to generate landscape images from doodles.

\section{Methodology}
The main challenge in producing high-quality synthetic fingerprints is \emph{estimating the probability distribution} of real fingerprints. 
Given the probability distribution, one can sample from this distribution to generate new fingerprints. 
However, obtaining such distribution is a very non-trivial task. 

\begin{figure}[ht]
\centering
\includegraphics[width=0.85\columnwidth]{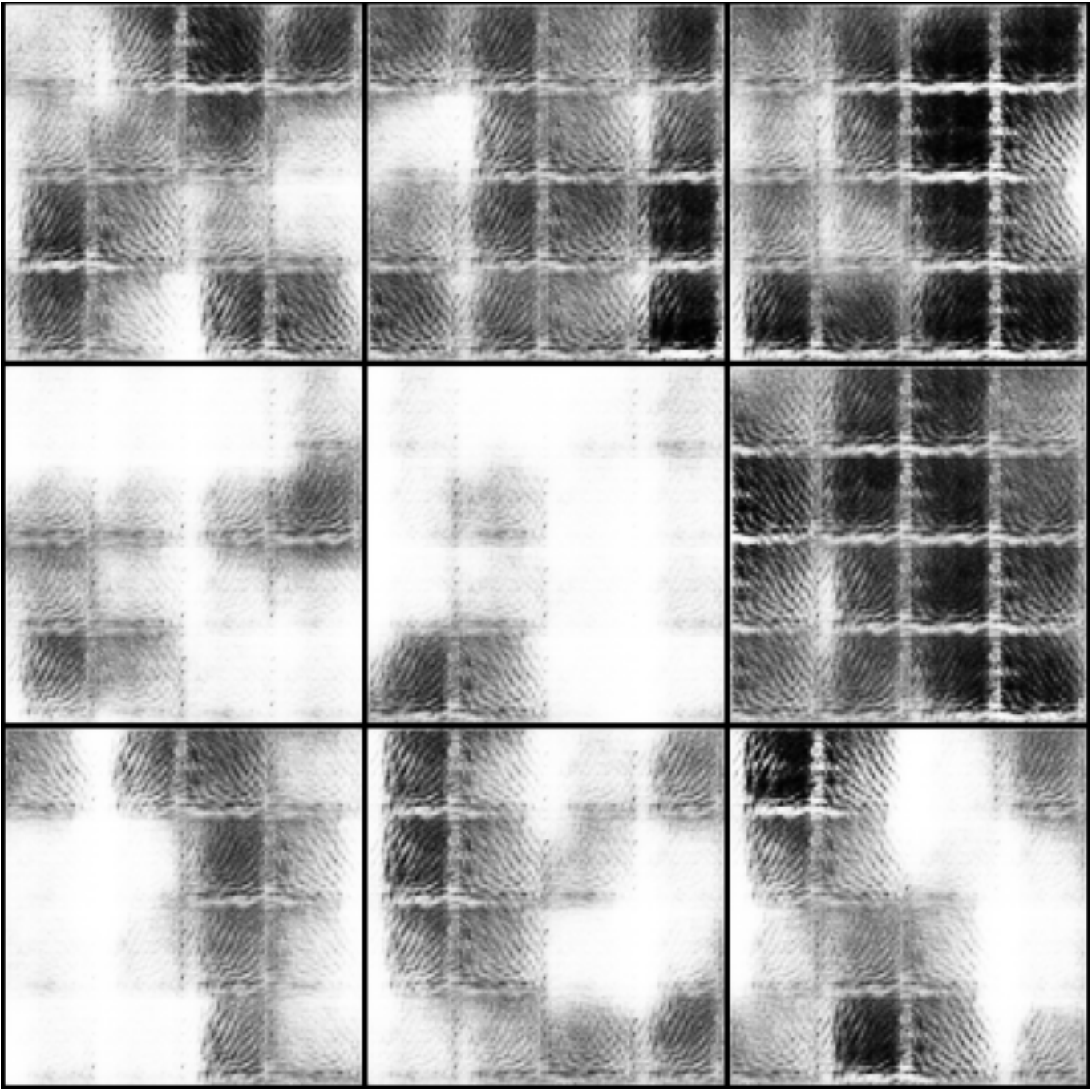}
\caption{The result of generating 256$\times$256 pixel images using GAN.}
\label{fig:gan256}
\end{figure}

In \sys{}, we rely on GANs to estimate the probability distribution of real fingerprints. 
Unfortunately, due to the small volume of the publicly available datasets, the GAN model cannot generalize well and produce realistic-looking samples. 
The prior art explores this approach. 
We also validate this idea and show that the generated samples by this approach are not acceptable (see~\fig{fig:gan256}).
As can be seen, these samples have deficient quality. They can easily be distinguished even without relying on sophisticated Machine Learning (ML) models. 
Therefore, in \sys{}, we capture the problem of generating synthetic fingerprints as a \emph{two-phase process}. 

In the first phase, we rely on a GAN model to estimate the probability distribution of real fingerprints and create a low-quality image out of a randomly generated vector representing the \emph{latent variable}. 
In the second phase, we train and use a \emph{Super-Resolution (SR)} model to transform the low-quality image into a realistic, high-quality sample.
In this phase, the details and texture of ridge endings and bifurcations within the fingerprints are embedded into the image.

\begin{figure*}[ht]
\centering
\includegraphics[width=\textwidth]{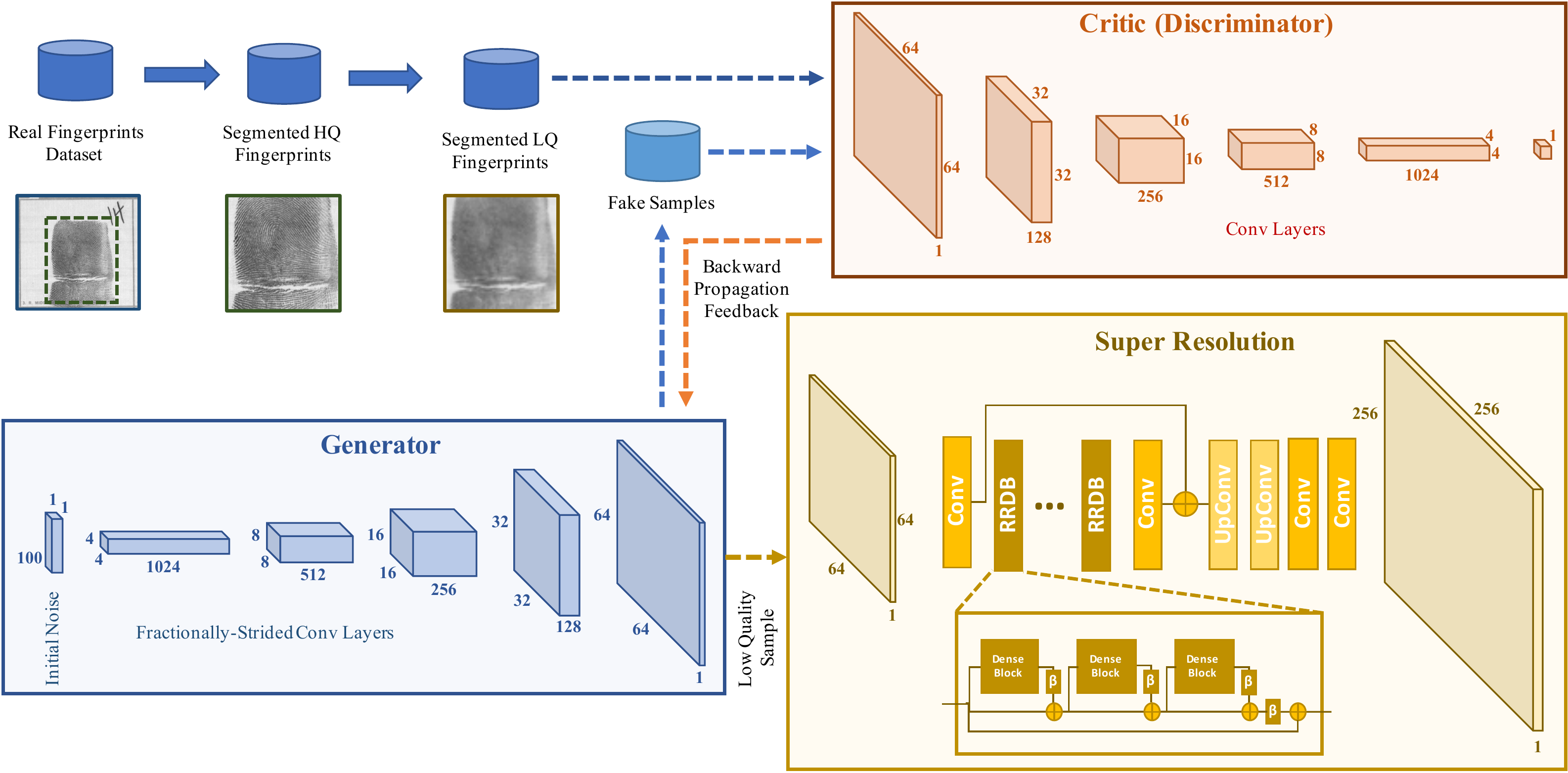}
\caption{Overall design of \sys{} for training (building the system) and execution (generating synthetic samples).}
\label{fig:end_to_end}
\end{figure*}

In order to train both GAN and SR models, we need a dataset of real fingerprints. 
In practice, however, these datasets comprise fingerprint images that are not centered and have unnecessary auxiliary information around the fingerprint image. 
Therefore, we need to preprocess the dataset to enhance the quality of the images produced by both models. 
\fig{fig:end_to_end} illustrates the overall design of \sys{} and the relationship between different components. 
In what follows, we describe each of these components in more detail.

\para{Pre-processing Phase: Fingerprint Segmentation}
As can be seen in~\fig{fig:end_to_end}, images in our dataset of real fingerprints contain some artifacts such as codes and numbers in the image.
Besides, fingerprints are not centered. 
NIST biometric image software (NBIS) provides specific tools for processing fingerprint images. 
In this phase, we have to detect the boundary of each fingerprint within each image, crop, and scale them accordingly. 
We observe that after this process, which we call \emph{segmentation} in this paper, the majority of the fingerprints have a resolution of 256$\times$256 pixels. 
However, the GAN training procedure relies on much lower resolution images. 
Therefore, we create a \textit{Low-Quality Database (LQD)} of 64$\times$64 images, which can be used to train GAN. 
The SR model, on the other hand, needs a \textit{High-Quality Database (HQD)} of 256$\times$256 images in addition to LQD. 

\para{Phase 1: Generating Synthetic Fingerprints using GAN}
After creating LQD, we train a GAN to generate (low-quality) synthetic fingerprints out of an input noise vector. 
There are dozen of different options for structures to use as our GAN network, in terms of the number and size of the layers and also the optimization loss and method used.
However, due to the small volume of the real fingerprints that are publicly-available, there are two main obstacles for the GAN model to converge: \emph{vanishing gradients} and \emph{mode collapse}.

In this paper, we choose Wasserstein GAN (WGAN)~\cite{arjovsky2017wasserstein} due to the following reasons. 
WGAN uses \emph{Earth-Mover distance} as its loss function for comparing the target and real distributions. 
Unlike the traditional Minimax loss function, the Earth-Mover distance is a true metric to measure distances in the space of probability distributions. 
This loss function helps the stabilization of the training process of GANs and reduces the possibility of several problems, including vanishing gradients and mode collapse.

The vanishing gradients problem arises because at the beginning of the training process, fake samples are easily distinguishable from the real samples; thus, the gradients computed during backpropagation are not helpful to tune the generative model.
In WGAN, however, the discriminator model outputs a number instead of a probability estimation.
The discriminator's job is to maximize the difference between the output number of real and fake samples (and not discriminate), thus, it is usually referred to as \emph{critic}. 
This enables the gradients to be informative, even at the beginning of the training process. 

The mode collapse problem is due to the fact that the generator can converge to a state that only produces a few plausible samples that can fool the discriminator. 
This problem is particularly important for us because our system has to have high entropy: generating many samples that are significantly different. 
Otherwise, \sys{} cannot scale to generating millions or billions of unique samples. 
Relying on WGAN helps us to avoid the mode collapse problem since the discriminator can separately be trained to optimality and quickly detect fake samples, forcing the generator to search for new samples. 

Even after incorporating the above optimizations and testing various configurations and different parameters, we observe that the trained GAN model is not capable of generating fingerprints with high quality for image sizes larger than 64$\times$64 pixels.
For instance, \fig{fig:gan256} shows the output of GAN for 256$\times$256 pixel samples.
In order to produce high-quality images similar to publicly available datasets (256$\times$256 pixel images), we need a second phase which we describe next.

\para{Phase 2: Generalization to High-Quality Images}
In the second phase of \sys{}, the low-quality image is transformed into a high-resolution image with a more detailed texture. 
Improving the resolution and quality of images is one of the challenging and interesting problems in the Computer Vision community. 
Traditional super-resolution mechanisms improve the quality of the input image using many but lower-quality images. 
However, in our case, the low-quality image is generated from an initial noise, and we cannot produce multiple low-quality images of the \emph{same} concept finger in Phase 1. 
Therefore, we have to explore \emph{single-image super-resolution} solutions that take as input only a single low-resolution image and produce a higher-quality image.
Single-image super-resolution is a significantly more challenging task.
Fortunately, GANs help in this regard too. 
Recent advances in this area include but are not limited to~\cite{ledig2017photo,wang2018sftgan,wang2018esrgan}.

After exploring several solutions in this area, we choose ESRGAN architecture with Residual-in-Residual Dense Blocks (RRDB). 
Details of our chosen architecture is provided in \fig{fig:end_to_end} and \sect{sec:exp}. 
At a high-level, the architecture consists of a series of RRDBs surrounded by convolutional (Conv) layers. 
There exist upsampling layers after RRDBs and before the convolution layers. 
In contrast to Phase 1, we remove Batch Normalization (BN) layers, which is shown to enhance the quality of the produced images~\cite{lim2017enhanced}. 
Adding BN layers increases the possibility of artifacts being added to the image. 
Relying on RRDB basic block enabled us to incorporate a higher number of hidden layers. 
We also leverage \emph{residual scaling}~\cite{wang2018esrgan}: residuals (the output of basic blocks) is scaled down by a constant ($0<\beta\leqslant1$) before being added to the main path. 
Additionally, we initialize the weights with small-variance random numbers to improve convergence. 

In our SR model, the discriminator differs from Phase 1 in which it is a Relativistic average Discriminator (RaD)~\cite{jolicoeur2018relativistic}. 
In Phase 1, the discriminator estimates the probability that a sample is real. 
In Phase 2, the discriminator estimates the probability that a real image is (relatively) more realistic than a fake one. More precisely, RaD is formulated as:
$$D(x_r,x_f)=\sigma(C(x_r))-\mathbb{E}_{x_f}[C(x_f)]$$
where $D(,)$ is the output of discriminator, $x_r$ is the real sample, $x_f$ is the fake sample, $C(.)$ is the non-transformed discriminator output, $\sigma(.)$ is the Sigmoid function, and $\mathbb{E}_{x_f}[.]$ represents taking an average over all fake samples in the batch. The discriminator's and generator's loss are: 
$$L_D=-\mathbb{E}_{x_r}[\text{log}(D(x_r, x_f))]-\mathbb{E}_{x_f}[\text{log}(1-D(x_f,x_r))]$$
$$L_G=-\mathbb{E}_{x_r}[\text{log}(1-D(x_r, x_f))]-\mathbb{E}_{x_f}[\text{log}(D(x_f,x_r))]$$
After training the SR model using the above loss functions, the model is used to generate the final synthetic fingerprints, as depicted in \fig{fig:end_to_end}. The hyperparameters of the training process are described in \sect{sec:exp}.

\section{Experimental Results}
\label{sec:exp}
In this section, we first provide the details of our Computational setup, our dataset of real fingerprints, the training procedure, followed by comparison with the prior art. In the end, we provide extensive analysis of the indistinguishability of the \sys{}'s synthetic fingerprints from real ones. 

\para{Computational Environment}
The experimental setup in which we train different components of \sys{} as well as synthetic fingerprint generation phases is a server equipped with $128$ GB of memory, two Intel Xeon E7 CPUs ($12$ core each), and four Nvidia Titan Xp GPUs (each with 12 GB of memory). 
We develop the DL components in Pytorch\footnote{Starting from the publicly available implementations at \\https://github.com/xinntao/BasicSR and\\ https://github.com/martinarjovsky/WassersteinGAN}. 

\begin{figure*}[ht]
\centering
\includegraphics[width=\textwidth]{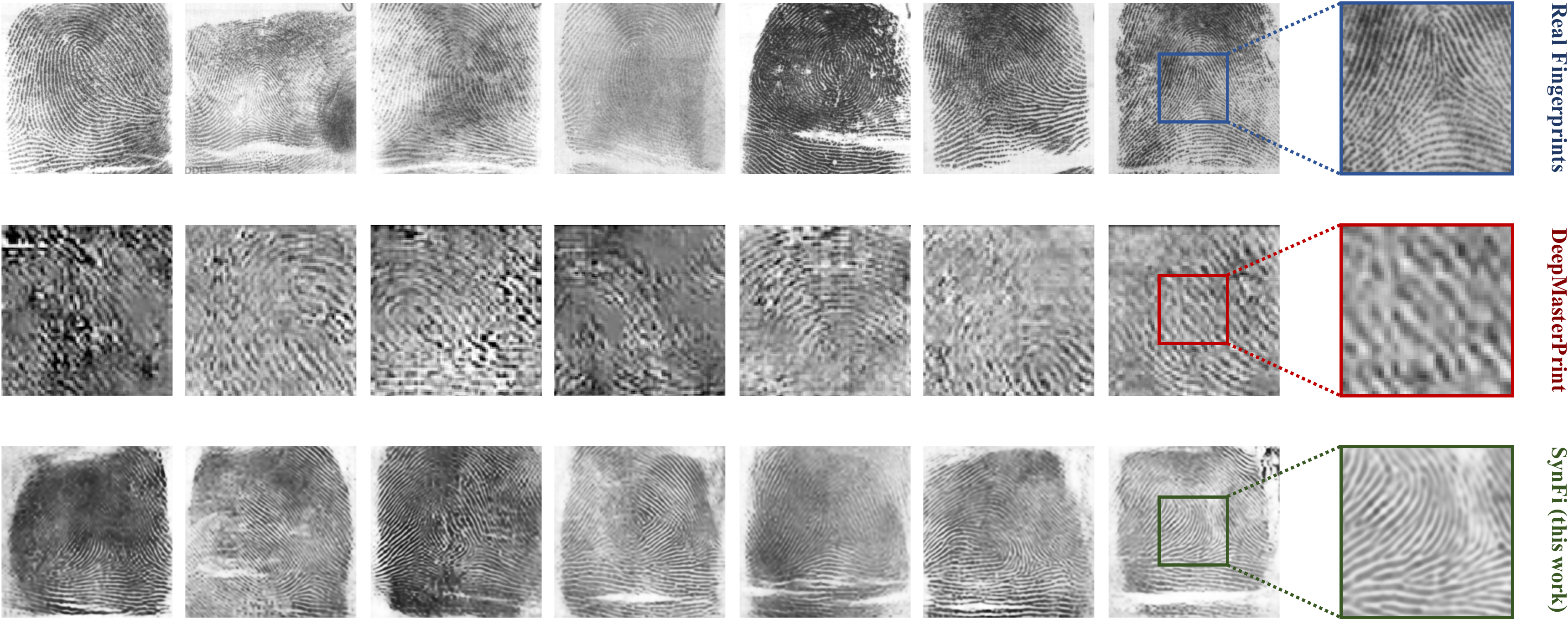}
\caption{Quality comparison between real and synthetic fingerprint samples.
\emph{Top row:} NIST dataset real fingerprint samples. 
\emph{Middle row:} Synthetic fingerprints generated by DeepMasterPrint~\cite{bontrager2018deepmasterprints}. 
\emph{Bottom row:} Synthetic fingerprints generated by \sys{} (this work).}
\label{fig:samples}
\end{figure*}
\para{Dataset}
Our dataset of real fingerprints is the one provided by National Institute of Standards and Technology (NIST) in 2009, named Special Dataset (SD09)\footnote{https://www.nist.gov/srd/nist-special-database-9}. 
We have used this dataset to train both major components of \sys{}: generative adversarial network and our super-resolution model. 
The NIST-SD09 dataset consists of 2700 subjects with all 10 fingerprint images.
There are two impressions of each finger, resulting in 54000 fingerprint images overall. 
The data format is 8-bit gray-scale \texttt{png} images. 
There are additional metadata associated with each fingerprint, including the subject gender and the NCIC class~\cite{komarinski2005automated}: arch (A), left-loop (L), right-loop (R), tented-arch (T), whorl (W), and scar or mutilation (S).
Most of the fingerprints belong to W, L, or R classes.

\para{Pre-processing Dataset}
We use the \texttt{nfseq} tool provided by the NIST Biometric Image Software (NBIS) package to pre-process the real fingerprint dataset.
As illustrated in \fig{fig:end_to_end}, this tool enables us to detect the precise boundary of the fingerprint and remove the unnecessary parts around the fingerprint itself. 
This step is crucial to enhance the quality of the images that are produced in both Phase 1 and Phase 2. 

\para{Architectures}
The generator and critic components of Phase 1 have similar architecture but in a reversed order. 
The generator starts with a noise vector of size $100$. 
Then the vector goes through a series of fractionally-strided convolutions in which the number of channels is reduced while the image size is increased, both by a factor of two. 
In the critic model, the intermediate layers are regular convolution layers. 

The SR model has a more complex architecture. 
In the beginning, there is a convolutional layer followed by a series of 23 basic blocks. 
In the end, there are two upsampling and two convolution layers. 
Each basic block consists of three residual sub-blocks where each sub-block has five convolutional layers. 
The convolutional layers have 64 channels with a kernel size of 3. 
The activation function in RRDB is a leaky ReLU with a slope of 0.2 in the negative part. 
The output of the SR model is a 256$\times$256 image with one channel (a gray-scale image).

\para{Synthetic Samples and Qualitative Comparison}
\fig{fig:samples} shows a set of samples of (i) real fingerprints in the NIST dataset, (ii) synthetic fingerprints generated by DeepMasterPrints~\cite{bontrager2018deepmasterprints}, state-of-the-art DL-based method, and (iii) synthetic samples generated by our system. 
As can be seen from this figure, the output of \sys{} is significantly more realistic compared to the prior art. 
Moreover, our methodology can generate a full impression of fingerprints as opposed to the partial fingerprints generated by DeepMasterPrints. 
The rightmost column shows a magnified view of the details of the impressions, which shows the quality of the produced samples in \sys{}. 
Next, we provide extensive experimental results to quantitatively compare \sys{} samples with real fingerprints.

\para{Indistinguishability and Quantitative Comparison}
As we briefly discussed before, one of the most important characteristics of synthetic fingerprints is their \emph{indistinguishability} from the real samples. 
Otherwise, not only synthetic samples cannot improve the quality and performance of authentication systems during development time, but also they cannot improve the security of storage systems for fingerprints as they can easily be distinguished and separated. 

\fig{fig:samples} shows that the synthetic fingerprints generated by our system are visually very similar to the baseline NIST dataset of real fingerprints.
However, to quantify how distinguishable synthetic fingerprints are from real ones, we perform the following analysis. 
We partition the subjects in the NIST dataset into training and test samples with 2200 and 500 subjects, respectively. Similarly, we create two disjoint sets of synthetic fingerprints, one for the training phase and one for the test phase. 
In order to minimize the classifier's bias, we put an equal number of real and synthetic fingerprints in the test dataset. 

We train six different machine learning models: a Logistic Regression (LR) model, a Support Vector Machine (SVM) with linear kernel, a Random Forest with 10 estimators, and three different Deep Neural Network (DNN) models with four, five, and eight layers. 
The training process is formalized as a binary classification problem in which real fingerprints are labeled as zero, and synthetic samples are labeled as one. 
After training the six ML models, we evaluate them on an unseen test set consisting of real and synthetic samples. 

The performance of these binary classifiers are reported in \tab{tab:indist} using three standard metrics: \emph{Accuracy (ACC)} which reflects the percentage of correct answers by the classifier, \emph{False Positive Rate (FPR)} which is defined as the ratio of wrongly classified samples as positive over all negative samples (both truly negative and incorrectly classified samples as positive). The third metric \emph{False Negative Rate (FNR)} is the ratio of the number of samples falsely labeled as negative over all positive samples (both true positives and misclassified samples as negative).  
During the training process, these ML models were trained to learn the underlying pattern of fingerprints and reached the training accuracy of up to 99.76\%.  
However, when evaluating these models on a set of unseen samples, the best performing classifier was the four-layer DNN with $100$ and $20$ neurons in the hidden layers and classification accuracy of \textbf{50.43\%}.
\textbf{In other words, the best classifier could distinguish synthetic fingerprints from real ones only 0.43\% better than a random guess.}

\begin{table}[H]
\centering
\caption{Analyzing indistinguishability of synthetic fingerprints against the real ones using various machine learning models.}
\label{tab:indist}
\resizebox{0.98\columnwidth}{!}{
\begin{tabular}{clrrr}\toprule
\textbf{Model Type} & \textbf{Model Description}                                                   & \multicolumn{1}{c}{\textbf{ACC(\%)}} & \multicolumn{1}{c}{\textbf{FPR(\%)}} & \multicolumn{1}{c}{\textbf{FNR(\%)}} \\\midrule
Logistic Regression & L2 regularization & 49.99 & \textbf{0.01} & 99.99 \\
Linear SVM & L2 regularization, C=1 & 50.01 & 0.06 & 99.91 \\
Random Forest & Using 10 estimators & 49.47 & 13.08 & 87.96 \\
4-Layer DNN & Hidden Layers: 100, 20 & \textbf{50.43} & 19.78 & 79.35 \\
5-Layer DNN & Hidden Layers: 100, 50, 10 & 50.35 & 11.36 & 87.93 \\
8-Layer DNN & \begin{tabular}[c]{@{}l@{}}Hidden Layers: \\800, 400, 200, 100, 50, 20 \end{tabular} & 49.85 & 23.25 & \textbf{77.03} \\
\bottomrule
\end{tabular}}
\end{table}

\begin{figure}[ht]
\centering
\includegraphics[width=0.98\columnwidth]{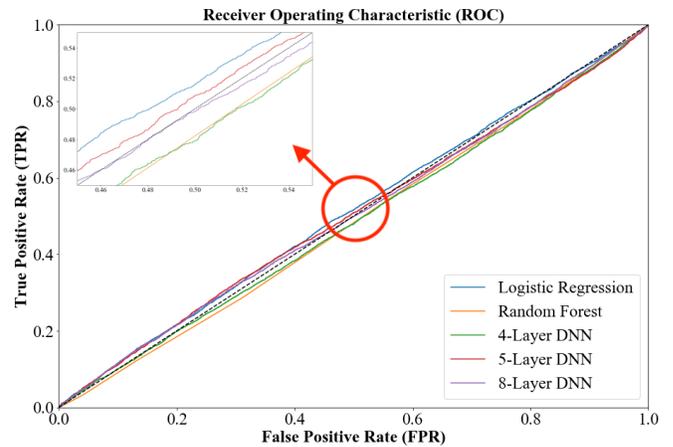}
\caption{The ROC curve of five different machine learning models in distinguishing real fingerprints from synthetic ones generated by \sys{}.}
\label{fig:roc}
\end{figure}

One can also analyze the effectiveness of the classifiers using a Receiver Operating Characteristic (ROC) curve.
The ROC diagaram depicts True Positive Rate (TPR) against FPR. TPR is defined as $\text{TPR}=1-\text{FNR}$.
\fig{fig:roc} shows the ROC curve of five ML models (ROC curve is not well-defined for SVMs). 
Relying on a purely random guess results in the black diagonal dashed line, which is the baseline for indistinguishability.  
Conceptually, ROC visualizes the fact that the classifiers' threshold for classifying an image as real or fake results in a trade-off between FPR and FNR (or TPR). 
Choosing a very low threshold leads to marking many real images as fake, hence, high FPR. 
Choosing a very high threshold results in outputting many fake images as real, thus, high FNR. 
However, as can be seen, regardless of the chosen value for the threshold, none of the classifiers can perform reasonably better than a purely random guess.

\section{Conclusion}
We present \sys{}, an automated framework to generate large volume of high-quality synthetic fingerprints. 
We formulate this task as two disjoint and parallel learning problems to cope with the limited availability of real fingerprint samples. 
Our fingerprint generation data flow involves two phases: one based on generative adversarial network and one based on super-resolution methodologies. 
We perform extensive experiments and empirically show that our synthetic fingerprints inherit fine-grained texture of real samples such as ridge endings and bifurcations.
Finally, we verify that the best performing machine learning model that we identified could distinguish synthetic fingerprints from real ones only 0.43\% better than a random guess, illustrating the effectiveness of \sys{} to enhance the security of fingerprint storage systems. 

\bibliographystyle{IEEEtran}
\bibliography{0_Main.bbl}

\end{document}